# An Experimental Quantum Bernoulli Factory

Raj B. Patel,[1,][*] Terry Rudolph,[2] and Geoff J. Pryde[1,][†]

[1]*Centre for Quantum Computation and Communication Technology and Centre for Quantum Dynamics, Griffith University, Brisbane 4111, Australia*
[2]*Department of Physics, Imperial College London, Prince Consort Road, London SW7 2AZ, UK*

(Dated: July 4, 2018)

There has been a concerted effort to identify problems computable with quantum technology which are intractable with classical technology or require far fewer resources to compute. Recently, randomness processing in a Bernoulli factory has been identified as one such task. Here, we report two quantum photonic implementations of a Bernoulli factory, one utilising quantum coherence and single-qubit measurements and the other which uses quantum coherence and entangling measurements of two qubits. We show that the former consumes three orders of magnitude fewer resources than the best known classical method, while entanglement offers a further five-fold reduction. These concepts may provide a means for quantum enhanced-performance in the simulation of stochastic processes and sampling tasks.

## INTRODUCTION

As the quantum information community continues its advance towards full-scale universal quantum computing[1, 2], along the way, a number of scenarios have been uncovered where quantum information processing offers a clear advantage over classical means. Furthermore, there exists certain tasks which are intractable using a classical computer but are made possible with quantum computing, supporting the notion of 'quantum supremacy'[3]. While there are examples where a quantum advantage may exist, unequivocal experimental proof is often unattainable.

Recently, the task of processing randomness—to transform probability distributions—has been identified by Dale *et al.* as a basic primitive for which quantum information processing offers advantages over classical stochastic techniques[4]. Specifically, the quantum information processing of randomness was shown to require fewer resources whilst also expanding the range of scenarios where such processing is possible. The randomness processing task has widespread applicability across science and is rooted in processes that are typically simulated by Markov chain Monte Carlo methods. Additionally, investigations in this area bear upon our fundamental understanding of quantum randomness[5]. In particular, they offer a new avenue for understanding the difference between epistemological classical randomness, owing to noncontextual ignorance about the real state of a system, and quantum randomness, for which no such interpretation is possible.

Here, we present quantum photonic experiments where polarisation qubits are used to encode sequences of random variables whose algorithmic processing yields quantum advantages in resource consumption. We show that quantum coherence reduces the consumption by several orders of magnitude compared to the best known classical method, whilst entanglement offers even further improvements.

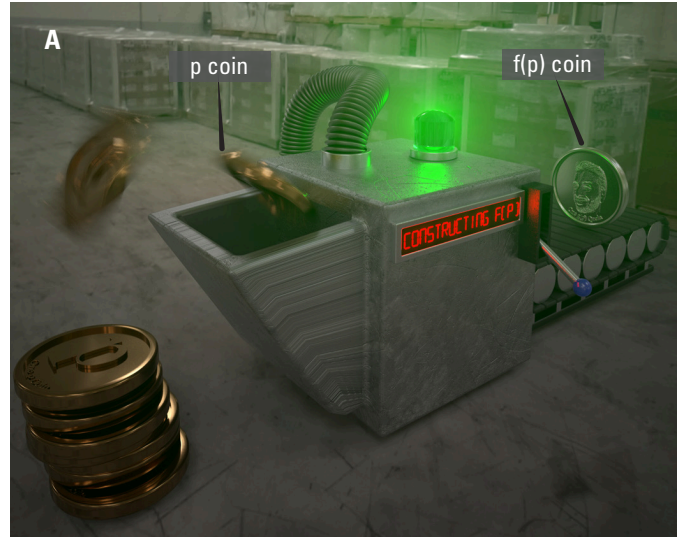

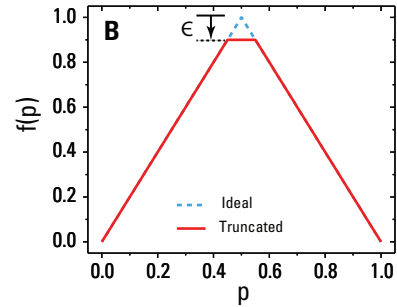

FIG. 1. (A) The concept of the Bernoulli factory. A sequence of iid coins, with an unknown bias $p$, are sampled and processed producing a new coin of bias $f(p)$. (B) The doubling function $f_\wedge(p) = 2p$, as per Eq. 2. The dashed blue plot shows the ideal function which cannot be constructed classically. The workaround is to truncate the function by $\epsilon$, shown by the solid red line.

Before describing the details our work, we set the scene with some simple examples.

Consider the scenario where a classical coin has an un-known bias $p$, which is the probability of a heads outcome from a coin toss. The challenge is then to simulate the behaviour of a fair coin, $f(p) = \frac{1}{2}$. von Neumann's solution[6] to this was to toss the coin twice, and provided $0 < p < 1$, if the outcomes are different output the value of the second coin toss, and if they are the same, then repeat. As a different example, suppose the task is to simulate the function $f(p) = p^2$ for $p \in [0,1]$. This can be achieved by tossing the coin twice and if each toss results in a head, output a head, otherwise output a tail. Indeed, some polynomials are well suited to this type of construction. While it is obvious that the function $f(p) = 2p(1-p)$ may be simulated by tossing a coin twice, the function $f(p) = 3p(1-p)$ requires noting that $3p(1-p) = 3p^2(1-p) + 3p(1-p)^2$[7], and as such, the coin must be tossed three times. In these examples, we have described the scenario of the so-called 'Bernoulli factory'[7–15], illustrated in Fig. 1A. Here, one can draw from a sequence of independent and identically distributed (iid) Bernoulli random variables (coins flips), i.e. $\mathbb{P}(X = 0 \equiv \text{Heads}) = p$ and $\mathbb{P}(X = 1 \equiv \text{Tails}) = 1-p$ for an unknown $p$, process the samples, and then output a new Bernoulli variable with success probability (or bias) $f(p) : (S \subseteq [0,1]) \to [0,1]$. These ideas were introduced by Asmussen et al.[16] in relation to the exact sampling of general regenerative processes, and later Keane and O'Brien[8] derived the necessary and sufficient conditions under which a Bernoulli factory exists for $f$. These conditions are i) $f$ must be continuous, ii) it must not approach 0 or 1 exponentially quickly, or reach 0 or 1 within its domain. We then have

$$\min\left(f(p), 1 - f(p)\right) \geq \min\left(p, 1-p\right)^k, \forall p \in S, \quad (1)$$

where $k \geqslant 1$.

## RESULTS

### The quantum Bernoulli Factory for $f(p) = 2p$

Recently, it was shown by Dale et. al.[4] that replacing the classical coin with a quantum coin or 'quoin' of the form $|p\rangle = \sqrt{p}|0\rangle + \sqrt{1-p}|1\rangle$, can yield some remarkable advantages. The extension to quoins enables algorithmic processing of coherent superpositions and entangled states, with a classical output. We will refer to this as the quantum Bernoulli factory (QBF) and the classical version as the CBF. One interesting feature of the QBF is that an advantage can be gained with quantum coherence alone. Furthermore, it was shown that the necessary and sufficient conditions in the quantum setting are now relaxed, allowing a larger class of functions to be constructed. In addition, functions constructable with both the CBF and the QBF were shown to require far fewer resources with the latter. Here, we report two photonic implementations of the QBF for the same function, one makes use of quantum coherence and entanglement while the other relies on quantum coherence only. We show that while both factories offer a quantum advantage, the use of entanglement offers a further improvement in performance over the best known CBF.

The function we choose to study, and perhaps the most important, is the 'Bernoulli doubling' function

$$f_\wedge(p) = 2p \equiv \begin{cases} 2p & p \in [0, 1/2] \\ 2(1-p) & p \in (1/2, 1] \end{cases}, \quad (2)$$

since it serves as a primitive for other factories[9]. That is, the ability to sample from this function allows any other analytical function to be constructed that is bounded at less than unity in $(0,1)$. Notice that this function cannot be constructed classically since $f_\wedge(0.5) = 1$ violates condition ii). In the classical setting, the workaround is to truncate the function by $\epsilon$ such that $f_\wedge(p) = 2p \cong \min(2p, 1-\epsilon)$[8, 9, 11–13], as shown in Fig. 1B. From ref. [4], the QBF for Eq. 2 can be realised by first rewriting the function as $f_\wedge(p) = 1 - \sqrt{1-4p(1-p)}$ and performing a series expansion,

$$f_\wedge(p) = \sum_{k=1}^{k_{\max}} \binom{2k}{k} \frac{1}{(2k-1)2^{2k}} (4p(1-p))^k$$
$$= \sum_{k=1}^{k_{\max}} q_k g_k \quad (3)$$

Here, $k > 0$, $q_k$ is independent of $p$, and $g_k = (4p(1-p))^k$. Typically, $k_{\max} = \infty$ however in realistic experimental scenarios, finite $k_{\max}$ values are considered. This representation allows us to reduce the problem to finding a construction for $g_k$, or $k$ consecutive heads outcomes of tossing a $g_1$-coin, where a $g_1$-coin is defined as a coin with a bias $g_1(p) = 4p(1-p)$. The main task is thus to efficiently produce such a $g_1$-coin. Performing a joint two-qubit measurement on two $p$-quoins $|p\rangle \otimes |p\rangle$ in the Bell-basis, $\{|\psi^\pm\rangle = (|01\rangle \pm |10\rangle)/\sqrt{2}, |\phi^\pm\rangle = (|00\rangle \pm |11\rangle)/\sqrt{2}\}$, we find that

$$\mathbb{P}\left(\phi^-|\psi^+ \cup \phi^-\right) = (2p-1)^2,$$
$$\mathbb{P}\left(\psi^+|\psi^+ \cup \phi^-\right) = 4p(1-p) = g_1(p),$$
$$\mathbb{P}\left(\phi^-|\psi^+ \cup \phi^-\right) + \mathbb{P}\left(\psi^+|\psi^+ \cup \phi^-\right) = 1. \quad (4)$$

The algorithm runs by first generating an index $k$ with a probability $q_k$. A joint measurement, in a restricted Bell-basis, is then performed on two $p$-quoins. If $k$ consecutive $|\psi^+\rangle$ outcomes are obtained, then the toss of an $f_\wedge(p)$-coin is heads, otherwise if the outcome of the measurement is $|\phi^-\rangle$, the output is tails.

### Two-qubit experimental QBF

The required measurements are well suited to our linear optics implementation shown in Fig. 2. A 404 nm,

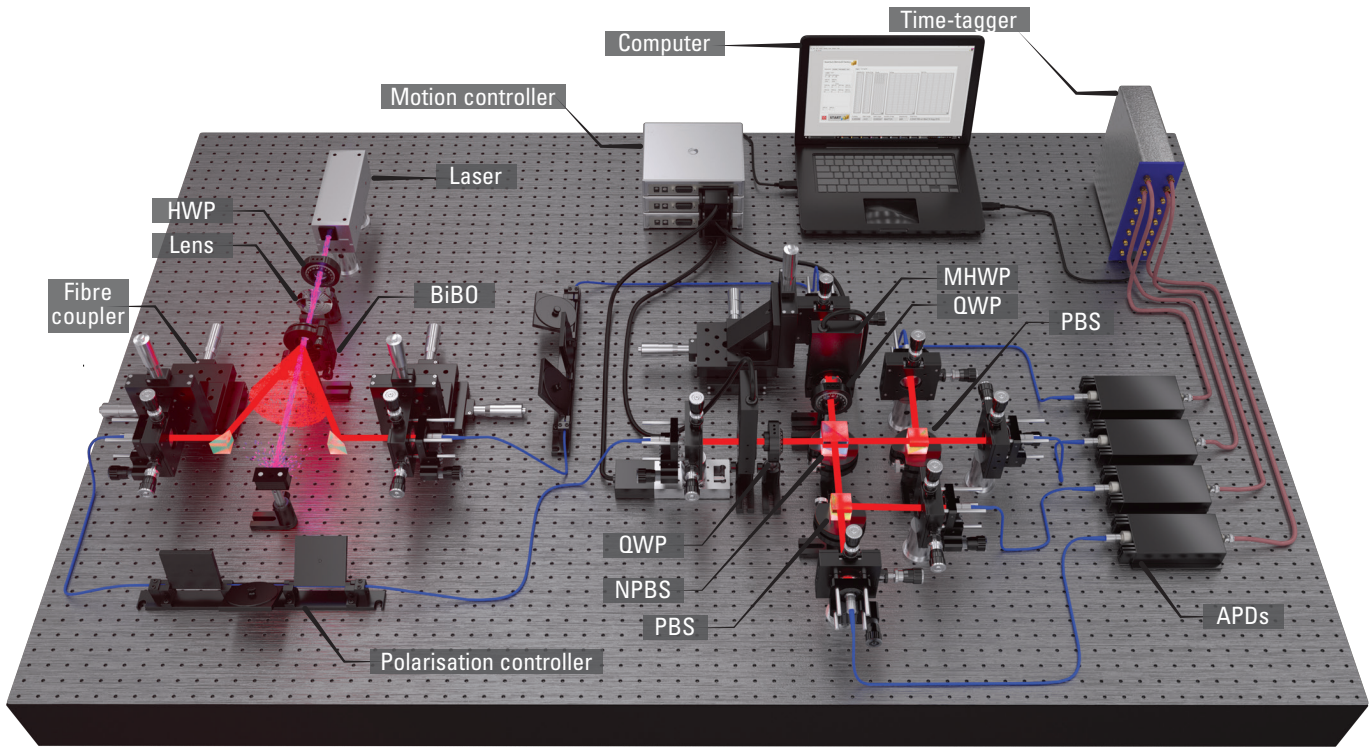

FIG. 2. Experimental arrangement for the QBF using joint measurements of two $p$-quoins. A pair of H-polarised photons are generated via type-I down-conversion in a non-linear BiBO crystal. They are sent to a Bell-state analyser arrangement containing additional motorised half-wave plates (MHWP) which set the bias value of each $p$-quoin, and quarter-wave plates (QWP), one at OA $+45°$ and one at OA $-45°$, which enables $|\psi^+\rangle$ and $|\phi^-\rangle$ to be identified. The photons interfere on a 50:50 non-polarising beamsplitter (NPBS) while polarising beamsplitters (PBS) enable H and V-polarised photons to be separated spatially before being detected using single-photon avalanche diodes (APDs). Detection events are time-tagged and analysed using a computer.

vertically (V) polarised, continuous-wave laser beam pumps a non-linear BiBO crystal generating a degenerate pair of horizontally (H) polarised photons, $|H\rangle_1 \otimes |H\rangle_2$. The photons are spectrally filtered using long-pass filters, and 3 nm wide band-pass filters centred at 808 nm, and are sent via single-mode fibre to a Bell-state analyser. This particular arrangement contains additional motorised half-wave plates (MHWP) which set the bias value of each $p$-quoin. It is well known that the standard linear optical Bell-state analyser[17, 18], relying on Hong-Ou-Mandel interference, is capable of unambiguously discriminating between the $|\psi^+\rangle$ and $|\psi^-\rangle$ Bell states. We implement an $X_{\frac{\pi}{2}} \otimes X_{-\frac{\pi}{2}}$ operation on the qubits before the measurement using quarter-wave plates (QWP) at optic axes (OA) $\pm 45°$ which allows the desired states, $|\psi^+\rangle$ and $|\phi^-\rangle$, to be identified. The photons interfere on a 50:50 non-polarising beamsplitter (NPBS) while polarising beamsplitters (PBS) enable H and V-polarised photons to be separated spatially before being detected using single-photon avalanche diodes (APDs). The sequence of detection events are time-tagged which allow us to identify the exact order in which a $g_1(p)$-coin toss resulted in a heads ($|\psi^+\rangle$) or tails ($|\phi^-\rangle$). The data is post-processed (see Methods for details) using a computer, and $f_\wedge(p)$ is constructed. Fig. 3A-D shows experimental data (circles) taken for $k_{\max} \in \{1, 10, 100, 2000\}$. We see that the data agrees strongly with the ideal theoretical plots (dotted lines). The red curves in each plot represent the expected data based on a model which takes into account the non-ideal splitting ratio of our NPBS, extinction ratios of our polarisation optics, and any mode-mismatch in our interferometer. The experimentally measured Hong-Ou-Mandel two-photon interference visibility was found to be $(99.7^{+0.3}_{-1.0})$ %. The experimental data shows an excellent agreement with our model. For lower values of $k$, the data shows a more rounded peak near $p = 0.5$ which becomes sharper for larger $k$. In our experimental run, we were able to generate up to a single $g_{2036}(p)$ coin, i.e. up to 2036 consecutive heads outcomes of the $g_1(p)$ coin. Higher order $g_k(p)$ coins are more susceptible to small experimental imperfections which may lead to erroneous coincident detections. For more reliable statistics, and for comparison later on, in Fig. 3D we restrict the expansion to $k_{\max} = 2000$ where we obtain $f_\wedge(0.5) = 0.935 \pm 0.003$. In Fig. 3E, we calculate the mean $p$-quoin consumption for each $f_\wedge(p)$-coin. Note that increasingly more quoins are required near $p = 0.5$, as we expect. We require an average (over $p$) of $\approx 11$ quoins to construct $f_\wedge(p) = 2p$ when utilising the quantum coherence and entangling measurements of two $p$-quoins.

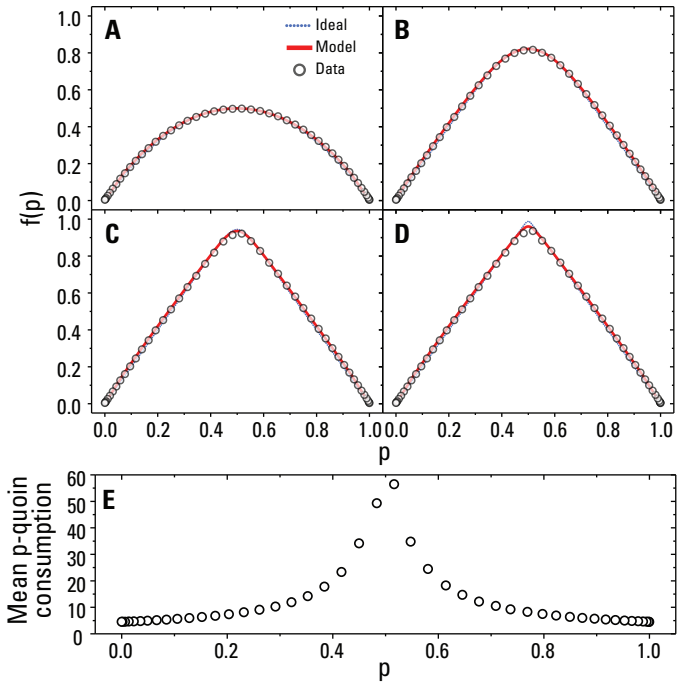

FIG. 3. Experimental data representing the construction of $f_\wedge(p) = 2p$, using joint measurements of two $p$-quoins for (A) $k_{\max} = 1$, (B) $k_{\max} = 10$, (C) $k_{\max} = 100$, (D) $k_{\max} = 2000$. The dotted blue lines are the ideal theoretical functions, and the red lines represents a model taking experimental imperfections into consideration. Error bars were too small to be included (see Methods). (E) Mean $p$-quoin consumption for $k_{\max} = 2000$.

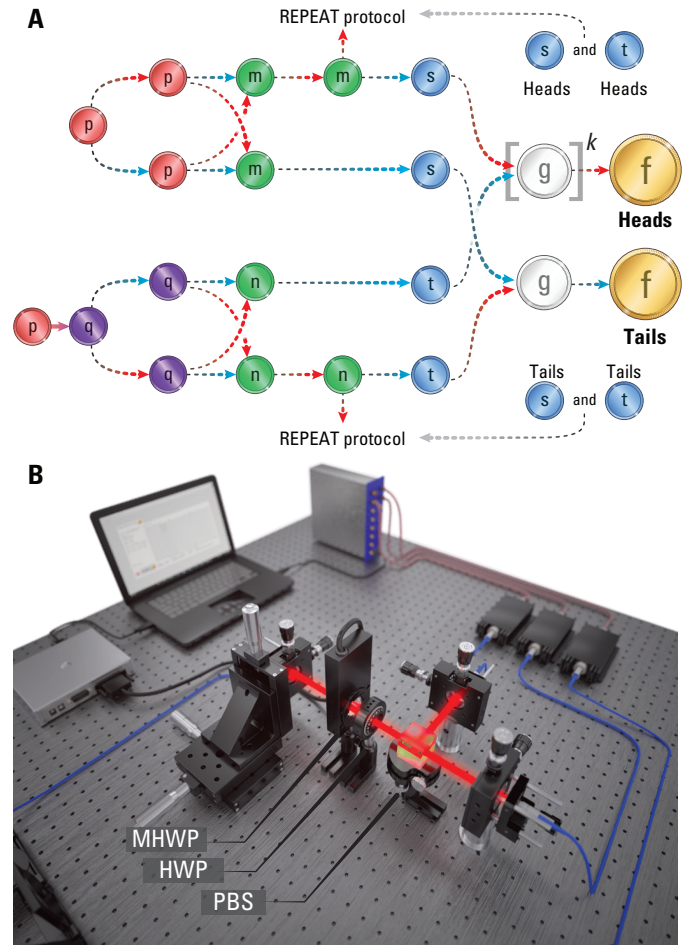

FIG. 4. Construction of $f_\wedge(p) = 2p$, using single-qubit measurements of $p$-quoins. (a) The algorithm we employ (see Supplementary Material). The upper (lower) branch begins with the measurement of the $p$-quoin in the $Z$-basis ($X$-basis). Red (blue) arrows indicate a heads (tails) outcome of the quoin toss. Failure to achieve the appropriate outcome requires the protocol to be repeated until success. (b) Experimental arrangement for the QBF using single-qubit measurements of $p$-quoins. A single photon from the source (not shown) is sent into the apparatus where it encounters a MHWP which sets $p$. A HWP set to optic axis enables $Z$-basis measurements to be performed for each $p$. The second photon, which serves as a herald, is detected directly by an APD. Setting the HWP to OA + 22.5° results in an $X$-basis measurement. Two sets of time-tag data are recorded allowing $p$ and $q$-quoins to be sampled.

### Single-qubit experimental QBF

We now show how $f_\wedge(p)$ can constructed using single-qubit measurements where we exploit quantum coherence alone. To do so, we employ the best known algorithm for constructing $g_1(p)$ with single-qubit measurements, which was recently demonstrated using superconducting qubits[19]. The algorithm makes use of additional intermediate quoins denoted by $q$, $m$, $n$, $s$, and $t$ each with a unique probability distribution. Fig. 4A illustrates the procedure where red (blue) arrows indicate a heads (tails) outcome. A more thorough description is provided in the Supplementary Material. To begin with, two $p$-quoins are produced, the second of which is measured in the $X$-basis to produce a $q$-quoin (lower branch). In the upper branch, a $p$-quoin is tossed twice, and if the outcome is different each time an $m$-quoin is produced with the outcome heads, otherwise tails is outputted. Similarly, in the lower branch where a $q$-quoin is tossed twice with different outcomes, an $n$-quoin with a heads outcome results. The $m$ and $n$-quoins are both tossed twice. In each case, if the first toss results in tails a new quoin is produced, $s$ or $t$, with a tails outcome. If however, the first toss gives heads and the second gives tails then heads is outputted in each case. Otherwise, the protocol is repeated from the beginning and two $p$-quoins are sampled again. Given the successful construction of an $s$ and $t$-quoin, if they have the value heads (tails) and tails (heads) respectively, the outcome of $g$-coin toss is heads (tails). If the outcome is the same each time, the protocol is repeated. From the successful sampling of the $g_1$-coin, $f_\wedge(p)$ can be constructed as outlined earlier.

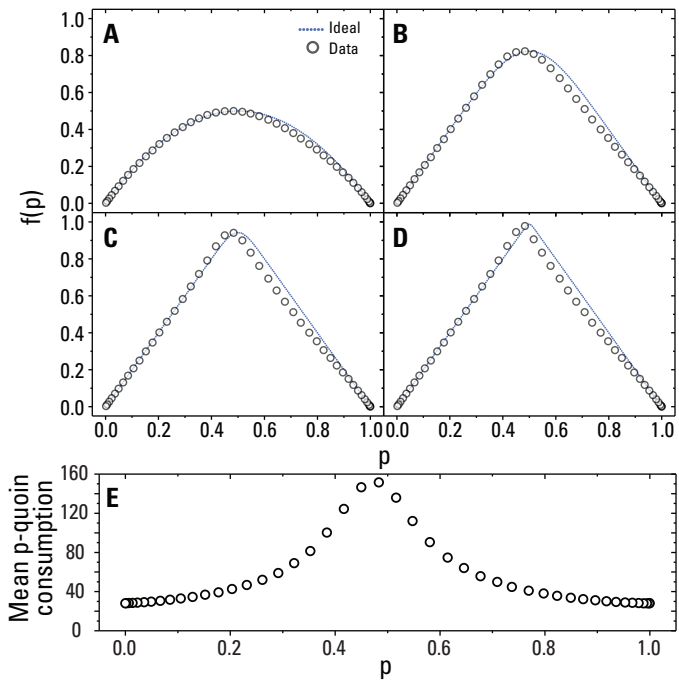

FIG. 5. Experimental data representing the construction of $f_\wedge(p) = 2p$, using single-qubit measurements of $p$-quoins for (A) $k_{max} = 1$, (B) $k_{max} = 10$, (C) $k_{max} = 100$, (D) $k_{max} = 2000$. The dotted blue lines are the ideal theoretical functions. Error bars were too small to be included (see Methods). (E) Mean $p$-quoin consumption for $k_{max} = 82$.

The experimental configuration is shown in Fig. 4B. Using the same photon-pair source as before, one photon is used as a herald while the other is sent to an arrangement of MHWP, HWP, and a PBS. Again, the MHWP sets the bias $p$ while the HWP set to OA (OA +22.5°) enables $Z$-basis ($X$-basis) measurements to be performed for each $p$. Time-tags are recorded for each measurement basis independently, the construction of $f_\wedge(p)$ then follows by sampling from the two data sets. Fig. 5A-D shows experimental data (circles) taken for $k_{max} \in \{1, 10, 100, 2000\}$. The data shows excellent agreement with theory under ideal conditions, albeit with a slight skew in the data which we attribute mechanical drift in the fibre coupling. As one might expect, single-qubit measurements, which do not rely on non-classical interference or multi-qubit coherence, can be performed with higher fidelity than joint measurements on two qubits. As such, in the case of $k_{max} = 2000$ we obtain $f_\wedge(0.5) = 0.977 \pm 0.006$.

Of particular interest, is a comparison of resource consumption between the two QBFs we have presented. For a fair comparison with the two-qubit QBF, we choose to restrict the series expansion of the single-qubit QBF to $k = 82$ which results in $f_\wedge(0.5) = 0.935 \pm 0.006$. Fig. 3E, shows the mean $p$-quoin consumption for each $f_\wedge(p)$-coin. Averaging over all $p$, we require $\approx 52$ quoins to construct $f_\wedge(p) = 2p$ when utilising the quantum coherence and single-qubit measurements of $p$-quoins, which is approximately a five-fold increase in resources over the two-qubit case.

### The quantum advantage

Owing to small experimental imperfections we are unable to exactly achieve $f_\wedge(0.5) = 1$, however, this does provide an avenue for comparing the QBF to the CBF. We can frame the situation as a refereed game played between two parties, the quantum player who has a QBF, and a classical player who has a CBF. The referee prepares $p$-quoins and sends them to the quantum player who is tasked with constructing, or approximating $f_\wedge(p) = 2p$, as best as they can. The quantum player can request a large, albeit, finite number of quoins. Their result is sent to the classical player who must reproduce it using a fewer resources. In the game, the quantum player achieves $f_\wedge(0.5) = 0.935$. The classical player's strategy is as follows. First, they perform a least-squares fit of the data using a positively weighted sum of Bernstein polynomials

$$\tilde{f}(p) = D \sum_{j=0}^{N} \frac{A_j}{D} \binom{N}{j} p^j (1-p)^{N-j} = D.H(p) \quad (5)$$

where $D = \sum_{j=0}^{N} A_j$ and $D.H(p) \leq 1 - \epsilon$. The parameters $A_j \geq 0$ are fitting parameters (see Supplementary Material) and $\epsilon = 1 - f_\wedge(0.5)$. As with the previously mentioned examples, polynomials of the form $p^j(1-p)^{N-j}$ require $N$ coins for exact sampling[14]. This approach takes into consideration the nuances of the experimental data which deviates from the ideal truncated function shown in Fig. 1B. It then follows from ref. [14] that the mean coin consumption is

$$\overline{N_c} \sim \frac{9.5DN}{\epsilon} \quad (6)$$

To determine the optimal $N$, the classical player performs an optimisation routine where the R-squared value is maximised for a range of $N$. For the data in Fig.3D (see Supplementary Material), $N = 27$, $D = 14.17$, and $\overline{N_c} \sim 56126$ coins on average which is three orders of magnitude greater than the quantum player, who wins the game. To the best of our knowledge, this is the optimum strategy that the classical player can employ.

Finally, we remark on how the resource consumption scales with $\epsilon$. From ref. [14] it was shown that the classical coin consumption for the truncated function shown in Fig. 1B is given by $\overline{N_c} \sim 19\epsilon^{-1}$. Taking into consideration the two-qubit QBF, we calculate the mean $p$-quoin consumption for a range of $\epsilon$. The two-qubit QBF presented here shows an improvement where the mean quoin consumption scales as $\overline{N_q} \sim 2\epsilon^{-0.5}$, which is in broad agreement with the scaling derived from the experimental data, $\overline{N_q} \sim 3\epsilon^{-0.4}$. As expected, this further supports the notion of a quantum advantage in resource consumption over the best known classical algorithm.

## DISCUSSION

The Bernoulli factory offers a fresh perspective from which information processing can be enhanced by quantum physics. Specifically, we have experimentally demonstrated a quantum advantage in the processing of randomness in a QBF under two different scenarios. Our work confirms that quantum coherence can provide a large reduction (three orders of magnitude) in resources over the CBF, and that quantum entanglement provides a further five-fold reduction. While our implementation utilises bipartite entanglement, an interesting question is how does this advantage scale when considering multipartite entangled systems? The QBF described here takes iid quoins as its input and outputs a coin. Lifting these restrictions, allowing quoins to be outputted rather than just coins, is expected to give rise to other classes of factories and constructible functions[21].

The QBF has also recently drawn comparisons to the quantum transducer[22], which is a model of an input-output process requiring a lesser amount of past knowledge and complexity compared to its classical counterpart to simulate the future state of the system. Further investigation is required to determine whether the QBF can offer additional insight in the study of processes which have a causal dependence.

### Acknowledgements

We acknowledge Theodore Yoder for useful discussions regarding the resource consumption in the CBF and QBF, as well as Sabine Wollmann for early contributions to the project. This work was supported by the Australian Research Council Centre of Excellence for Quantum Computation and Communication Technology (Project numbers CE110001027 and CE170100012) and the Engineering & Physical Sciences Research Council.

### Author contributions

R.B.P. and G.J.P. designed the experiments and R.B.P. performed them. T.R. provided theory support. R.B.P. analysed the data with input from T.R. and G.J.P. The manuscript was written by R.B.P. with input from the other authors.

## MATERIALS AND METHODS

### Data Processing

For each setting of $p$, $\sim 3 \times 10^6$ time-tags are recorded. Since two photons are always created at the same time during the down-conversion process, the data is filtered such that only tags occurring within a window of 6.25 ns remain. This process eliminates most of the spurious events due to ambient light or detector dark counts. The data is traversed and a tally is kept for the number of coincidence detections corresponding to a heads outcome for each of the $g_k(p) \equiv g^k(p)$ coins as well as the number of tails outcomes. $g_k(p)$, is then calculated as $\#\text{heads}_k/(\#\text{heads}_k + \#\text{tails})$. We also note that a tally is kept of the total number $p$-quoins required to produce an $f_\wedge(p) = 2p$ coin, which also includes cases where coincident detections correspond to neither a head nor a tail $g_1(p)$-coin. These events may occur due to imperfect extinction of the polarisation optics or the finite polarisation dependence of the non-polarising optics. This total is then weighted by $q_k$ allowing the mean $p$-quoin consumption to be calculated.

Poissonian uncertainties arise because we count a large number of $g_k$-coins within a fixed data collection time window. Errors quoted throughout the main text were calculated assuming Poissonian counting statistics of the coincidence detections which are integrated to give $\#\text{heads}_k$ and $\#\text{tails}$. Errors in $f_\wedge(p)$ typically varied between $\pm 2 \times 10^{-4}$ and $\pm 3 \times 10^{-3}$.

---

* r.patel@griffith.edu.au
† g.pryde@griffith.edu.au
[1] T. D. Ladd, F. Jelezko, R. Laflamme, Y. Nakamura, C. Monroe, and J. L. O'Brien, Nature **464**, 45 (2010).
[2] R. B. Patel, J. Ho, F. Ferreyrol, T. C. Ralph, and G. J. Pryde, Science advances **2**, e1501531 (2016).
[3] A. W. Harrow and A. Montanaro, Nature **549**, 203 (2017).
[4] H. Dale, D. Jennings, and T. Rudolph, Nat Commun **6**, 8203 (2015).
[5] P. Grangier and A. Auffèves, arXiv:1804.04807 (2018).
[6] J. von Neumann, Appl. Math Ser **12**, 36 (1951).
[7] J. Wästlund, *Functions arising by coin flipping*, Tech. Rep. (Technical Report, KTH, Stockholm, 1999).
[8] M. S. Keane and G. L. O'Brien, ACM Trans. Model. Comput. Simul. **4**, 213 (1994).
[9] Ş. Nacu and Y. Peres, Annal. Appl. Probab. **15**, 93 (2005).
[10] E. Mossel, Y. Peres, and C. Hillar, Combinatorica **25**, 707 (2005).
[11] K. Łatuszyński, I. Kosmidis, O. Papaspiliopoulos, and

G. O. Roberts, Random Structures & Algorithms **38**, 441 (2011).

[12] A. C. Thomas and J. H. Blanchet, arXiv preprint arXiv:1106.2508 (2011).

[13] J. M. Flegal and R. Herbei, Electronic Journal of Statistics **6**, 10 (2012).

[14] M. Huber, Combinatorics, Probability and Computing **25**, 577 (2016).

[15] M. Huber, Methodology and Computing in Applied Probability **19**, 631 (2017).

[16] S. Asmussen, P. W. Glynn, and H. Thorisson, ACM Trans. Model. Comput. Simul. **2**, 130 (1992).

[17] K. Mattle, H. Weinfurter, P. G. Kwiat, and A. Zeilinger, Phys. Rev. Lett. **76**, 4656 (1996).

[18] D. Bouwmeester, J.-W. Pan, K. Mattle, M. Eibl, H. Weinfurter, and A. Zeilinger, Nature **390**, 575 (1997).

[19] X. Yuan, K. Liu, Y. Xu, W. Wang, Y. Ma, F. Zhang, Z. Yan, R. Vijay, L. Sun, and X. Ma, Phys. Rev. Lett. **117**, 010502 (2016).

[20] T. Yoder, https://www.scottaaronson.com/6s899/tedyoder.pdf (2015).

[21] J. Jiang, J. Zhang, and X. Sun, Phys. Rev. A **97**, 032303 (2018).

[22] J. Thompson, A. J. P. Garner, V. Vedral, and M. Gu, npj Quantum Inf. **3**, 6 (2017).

# Supplementary Material: An Experimental Quantum Bernoulli Factory

Raj B. Patel,[1,*] Terry Rudolph,[2]
& Geoff J. Pryde,[1,*]

[1]CQC2T and Centre for Quantum Dynamics, Griffith University,
Brisbane 4111, Australia

[2] Department of Physics, Imperial College London, Prince Consort Road, London SW7 2AZ, UK

[*]To whom correspondence should be addressed; E-mail: r.patel@griffith.edu.au
or g.pryde@griffith.edu.au

## S1. Constructing $g_1(p)$ in the single-qubit QBF

Following the treatment in ref. (*1*), here we detail the construction of the $g_1(p)$-quoin in the single-qubit QBF, as illustrated in Fig. 4A in the main text. First, we begin with two $p$-quoins, then second of can be measured in the $X$-basis produces a $q$-quoin with an outcome

$$\mathbb{P}_q(\text{Heads}) \;=\; 1 + 2\sqrt{p(1-p)}. \tag{S1}$$

The $p$-quoin is tossed twice (upper branch) to generate a virtual $m$-quoin. Two different (identical) outcomes leads to a toss of the $m$-quoin with a value of heads (tails) with probability

$$\begin{aligned}
\mathbb{P}_m(\text{Heads}) \;&=\; \mathbb{P}_p(\text{Heads})\mathbb{P}_p(\text{Tails}) \\
&+\; \mathbb{P}_p(\text{Tails})\mathbb{P}_p(\text{Heads}) \\
&=\; 2p(1-p).
\end{aligned} \tag{S2}$$

Similarly an $n$-quoin from two tosses of a $q$-quoin in the same manner giving

$$\begin{aligned}\mathbb{P}_n(\text{Heads}) &= \mathbb{P}_q(\text{Heads})\mathbb{P}_q(\text{Tails}) \\ &+ \mathbb{P}_q(\text{Tails})\mathbb{P}_q(\text{Heads}) \\ &= 1/2 - 2p(1-p).\end{aligned} \quad (S3)$$

The next step is to toss the $m$-quoin ($n$-quoin) twice if the first toss results in tails a we produce an $s$-quoin ($t$-quoin), with a tails outcome. If however, the first toss gives heads and the second gives tails then heads is outputted

$$\begin{aligned}\mathbb{P}_s(\text{Heads}) &= \mathbb{P}_m(\text{Heads})\mathbb{P}_m(\text{Tails}) \\ &+ \mathbb{P}_m(\text{Heads})(1-\mathbb{P}_m(\text{Tails})\mathbb{P}_s(\text{Heads}), \\ \Rightarrow \mathbb{P}_s(\text{Heads}) &= \frac{m}{(1+m)}, \text{ and} \quad (S4) \\ \mathbb{P}_t(\text{Heads}) &= \frac{n}{(1+n)}. \quad (S5)\end{aligned}$$

Otherwise, the protocol is repeated. An $s$ and $t$-quoin are tossed, if the result is heads (tails) and tails (heads), respectively, the outcome of $g$-quoin toss is heads (tails) with probability

$$\begin{aligned}g_1(p) &\equiv \mathbb{P}_g(\text{Heads}) = \mathbb{P}_s(\text{Heads})\mathbb{P}_t(\text{Tails}) \\ &+ [1-\mathbb{P}_s(\text{Heads})\mathbb{P}_t(\text{Tails}) \\ &- \mathbb{P}_s(\text{Tails})\mathbb{P}_t(\text{Heads})]\mathbb{P}_g(\text{Heads}) \\ &= 4p(1-p). \quad (S6)\end{aligned}$$

else, the protocol is repeated.

## S2. Bernstein polynomial fit of the data

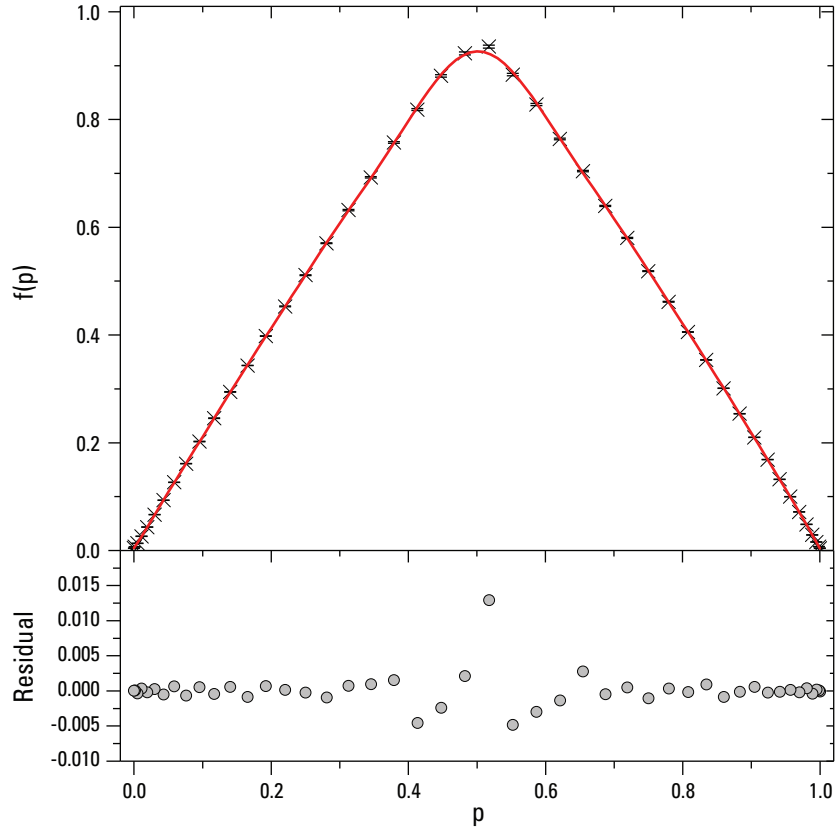

**Fig. S1.** Least squares fit of $f_\wedge(p) = 2p$. For the $k = 2000$ of the two-qubit QBF, the experimental data was fitted using a sum of Bernstein polynomials given by Eq. 5 in the main text. The fit presented here for order $N = 27$ (and corresponding R-squared value of 0.999992) was determined by maximising the R-squared value for a range of $N$. The fit was weighted by the error bars shown, which were calculated assuming Poissonian statistics.

# References

1. X. Yuan, *et al.*, *Phys. Rev. Lett.* **117**, 010502 (2016).